\documentclass[10pt, conference, compsocconf]{IEEEtran}

\usepackage{amsmath}
\usepackage{graphicx,psfrag,epsf}
\usepackage{enumerate}
\usepackage{url} % not crucial 
\usepackage{framed}
\usepackage{caption}
\usepackage[usenames, dvipsnames]{color}
\usepackage[table]{xcolor}
\usepackage{array}
\usepackage{booktabs}

\begin{document}
\title{Safely and Quickly Deploying New Features with a Staged Rollout Framework Using Sequential Test and Adaptive Experimental Design}

\author{\IEEEauthorblockN{Zhenyu Zhao\IEEEauthorrefmark{1}, Mandie Liu\IEEEauthorrefmark{2}, and Anirban Deb\IEEEauthorrefmark{3}}
\IEEEauthorblockA{Uber Technologies, Inc.\\
San Francisco, USA\\
Email: \IEEEauthorrefmark{1}
zhenyuz@uber.com \\
\IEEEauthorrefmark{2}
mandie@uber.com  \\
\IEEEauthorrefmark{3}
anirban@uber.com }
}




\maketitle

\begin{abstract}
During the rapid development cycle for Internet products (websites and mobile apps), new features are developed and rolled out to users constantly. Features with code defects or design flaws can cause outages and significant degradation of user experience. The traditional method of code review and change management can be time-consuming and error-prone. In order to make the feature rollout process safe and fast, this paper proposes a methodology for rolling out features in an automated way using an adaptive experimental design. 
Under this framework, a feature is gradually ramped up from a small proportion of users to a larger population based on real-time evaluation of the performance of important metrics. If there are any regression detected during the ramp-up step, the ramp-up process stops and the feature developer is alerted.
There are two main algorithm components powering this framework: 1)  a continuous monitoring algorithm -  using a variant of the sequential probability ratio test (SPRT) to monitor the feature performance metrics and alert feature developers when a metric degradation is detected, 2) an automated ramp-up algorithm - deciding when and how to ramp up to the next stage with larger sample size. This paper presents one monitoring algorithm and three ramping up algorithms including time-based, power-based, and risk-based (a Bayesian approach) schedules. These algorithms are evaluated and compared on both simulated data and real data. 
There are three benefits provided by this framework for feature rollout: 1) for defective features, it can detect the regression early and reduce negative effect, 2) for healthy features, it rolls out the feature quickly, 3) it reduces the need for manual intervention via the automation of the feature rollout process. 
\end{abstract}


\section{Introduction} \label{intro}
During web and mobile software product development cycles, innovations and improvements are continuously made to the products by developing and rolling out new features. Integration of features to the mobile app or digital website requires changes to the code base on the client or server sides. In a complex world of micro-services and intra-service dependencies, it's very hard to ensure reliability in a manual way. While new features are designed and tested cautiously before being released to end users, it's inevitable that a small fraction of these features will have bugs in the code or flaws in their design. Such problematic features will degrade the user experience (e.g. causing application crash, slowing down the performance, draining mobile battery, blocking users from requesting a trip, etc.) when they are released. 

Improper ways to release a new feature include (1) making the feature available to all users at once; (2) releasing a feature without a proper monitoring and alerting system in place.
Due to the uncertainty of impact embedded in the new features, releasing the feature through an unguided process can be risky:
\begin{itemize}
\item The negative impact may go unnoticed for a long time resulting in a bad customer experience.
\item The problematic feature may impact a large number of users, even if it is reverted timely.
\item When a metric degradation is detected, it is hard to attribute the degradation to a specific feature and fix a bug.
\end{itemize}

Online controlled experimentation, also known as A/B testing, is one way to evaluate a feature's impact before formal release \cite{Keppel1992, KOHAVI2007, KOHAVI2009, KOHAVI2013, KOHAVI2014, Tang2014recsys, Frasca2014recsys, Smallwood2014recsys, zhao2016online, zhao2017segment}. 
In a standard A/B test, users are split into treatment group (with the new feature turned on) and control group (with the new feature turned off), and the analysis is performed on the user engagement and business metrics when the experiment concludes. 
While experimentation can help detect potential flaws in the feature before its formal release, it is worth noting that such feature release process can still be risky post experimentation. If the experiment was run on certain population (let's say in couple of cities) and then rolled out to a population that were not tested against (rolled out to all cities) then the post-experiment release can be risky since it may affect an under-represented population in the experiment. Thus, even with the presence of an A/B test, a proper rollout process is needed for the post experimentation feature rollout. 

A reliable feature release process can systematically reduce the negative impact of problematic feature updates. A reliable feature release process has following desirable properties: 1) early detection of regression, 2) reverting feature rollout if regression is detected, 3) reducing the impact scope in terms of user size, 4) rolling out flawless features at a fast speed, and 5) requiring minimum human attention and intervention in the process.

Based on the criteria mentioned above, this paper proposes a framework for rolling out features autonomously using an adaptive experimental design. The rollout starts with a small proportion of users getting exposed to this feature, gradually ramps up to a larger user population, and eventually rolls out to the entire target population. 
This rollout framework is composed of two main algorithmic components making the process both intelligent and autonomous. First, a continuous monitoring algorithm based on sequential test that detects regressions early so that the team can be alerted and the feature can be reverted in a timely manner. Second, an adaptive ramp-up algorithm determining when and how much  to ramp up to based on required sample size to get enough power at the end stage.
This proposed framework satisfies the five properties for a safe feature rollout process. It enables engineers to quickly evaluate and roll out new features with a controlled amount of risk, without too much manual work and intervention during the process. As a result, this framework powers the development and rollout of mobile and web software products. 

The contribution of this paper includes:
\begin{itemize}
\item Proposing a staged rollout framework for releasing new features autonomously by using a continuous monitoring algorithm and a ramp-up algorithm.
\item Introducing a scalable nonparametric variance estimator to be used with the sequential test to adjust the sample correlation when the observation is not independent.
\item Providing formulas for estimating power and required sample size for the sequential test.
\item Proposing and compared three ramp-up algorithms: time-based, power-based, and risk-based.
\item Evaluating the empirical performance of the staged rollout framework through examples.  
\end{itemize}

This paper is organized as follows:  
Section~\ref{sec:background} introduces the background and terminology of this work, as well as its relation to standard A/B testing; 
Section ~\ref{sec:staged-rollout} presents the concept of the staged rollout framework for feature release; 
Section ~\ref{sec:monitoring} describes the monitoring algorithm: sequential probability ratio test with nonparametric variance estimation; 
Section ~\ref{sec:rampup} introduces three ramp-up algorithms: the time-based ramp-up, the power-based ramp-up based on statistical power of sequential test, and the risk-based ramp-up algorithm based on Bayesian methods;
%Section ~\ref{sec:autonomous-rollout} illustrates the autonomous rollout framework by combining different components together.
Section ~\ref{sec:example} evaluates how this rollout framework works in practice through both real data and synthetic data examples;
and Section ~\ref{sec:discussion} summarize the work and discuss practical considerations.

\section{Background} \label{sec:background}
\subsection{Terminology}
In this section, the terminology is defined and introduced for this paper. Some of the terminologies are commonly known in the software development world, and some of them have specific meanings and scopes in this paper.
\begin{itemize} 
\item {\bf Feature} A new feature is defined as any change to the product that involves a code or configuaration change. The change can happen either in the client codebase, or in the backend codebase on the server. A feature can be as large as an app redesign, and as small as a few lines of code or configuaration change that are not visible to users. 
\item {\bf Feature Flag}  Feature flags are a gating system to control code flow in the product. A feature flag can turn on and off a block of code (as a feature). In other words, a feature flag can  be used to control if a user can see a new feature or not.  For example, turning on a feature is controlled by setting the feature flag value to True, while a feature can be turned off by setting the value to False. 
\item {\bf Rollout} The process of exposing a new feature to the user population is defined as a rollout. The rollout starts with the state of no users getting exposed to the new feature and ends with the feature turned on for all users. 
\item {\bf Staged Rollout} The staged rollout framework proposed in this paper breaks the feature rollout process into different stages, each stage with a different proportion of the user population getting exposed to the feature. In this framework, a feature is gradually rolled out to the user population from smaller sample sizes to larger sample sizes.  
\item {\bf Revert} Reverting a feature means turning off the feature for all users. When the new feature is turned off, a default experience (often the legacy version of the feature) is served to the user. 
\item {\bf Ramp-up} Ramping up a feature means turning on the feature for more users. In the staged rollout framework, each subsequent stage has a larger sample size (in proportion) than the previous stage, thus moving forward one stage is equivalent to ramping up the feature to more users. 
\item {\bf Regression} A regression is a situation where a feature stops functioning as intended or performs in a suboptimal manner. A regression can cause user experience degradation. A regression can be caused by a new feature that contains a code bug or design flaw.
\item {\bf Outage} An outage, also known as downtime, is a serious regression situation that the core function or the entire app stops functioning. A problematic feature can cause an outage. 
\end{itemize}

\subsection{Relation to Standard A/B Testing}

\begin {table*}
\caption {Difference between Staged Rollout and Standard A/B Testing}
\label{tab:difference_experiment}
{\small
\begin{tabular}{ | l | l | l | }
\hline
& {\bf Staged Rollout} & {\bf Standard A/B Testing} \\
\hline
Main Purpose & Automated Feature Release Process & Feature Evaluation Process \\
\hline
Question Trying to Answer  & If the feature is causing a regression & If the feature is successful \\
\hline
Experiment Design and Configuration  & Multiple Adaptive Stages  & One Fixed Stage \\
\hline
Metrics  & Core user engagement and software performance  & Feature related user engagement \\
\hline
Statistical Test  & Sequential Test  & Fixed Horizon Test \\
\hline
Analysis Frequency  & Continuous Monitoring  & One-Time Fixed Horizon \\
\hline
\end{tabular}
}
\end{table*}

There are similarities between the staged rollouts and standard A/B testing. In both cases, there is a randomized control group to compare with the treatment group. A large portion of the infrastructure can also be shared between these two frameworks. 
However, there are major differences between them two, and the standard A/B testing cannot solve the problem that the staged rollouts are focused on. See Table~\ref{tab:difference_experiment} for a list of difference between staged rollouts and standard A/B testing. In practice, the staged rollout framework can be used by itself to roll out a feature, but it can also be used within an A/B testing to monitor and ramp up the experiment.

\section{Staged Rollout Framework Overview} \label{sec:staged-rollout}
The design for the staged rollout framework comes with two main components. 

First, the feature rollout process is staged such that the feature is gradually ramped up. The initial stage comes with a relatively small sample size (e.g. $1\%$ users in treatment), while the last stage consists of a large sample size reflecting the final rollout goal, usually $100\%$ of users in the treatment. 
The stages are designed to be increasing in sample size. The number of stages and the sample size at each stage can be determined by either manual inputs or algorithms (to be introduced in following sections). 

Second, the feature rollout process is monitored and controlled by statistical algorithms to evaluate the feature impact and infer the optimal rollout actions: revert, ramp up, or stay on the current stage. There are two main statistical algorithms supporting the rollout decision-making: a continuous monitoring algorithm and an adaptive ramp-up algorithm. 
The monitoring algorithm continuously monitors the metrics for the rollout via a sequential test (Section \ref{sec:monitoring}). If there is no significant difference detected in the metrics between treatment and control, the rollout process will continue. On the other hand, if there is a significant difference detected, an alert with the metric signal will be sent to the feature developers and the rollout can be reverted by the developer or the system. In addition to the monitoring algorithm for deciding when to revert a rollout, there are three methods to decide when to ramp up to the next stage introduced in Section \ref{sec:rampup}. The time-base ramp-up schedule is the most basic but transparent approach. The power-based rampup aims to achieve sufficient statistical power at a pre-set sensitivity level (minimum detectable difference) within a given time frame. The risk-based ramp-up algorithm sets the rollout speed and scope according to the signals collected and risk tolerance based on a Bayesian framework.  

Table \ref{tab:staged_rollout_template} shows an example scenario for a staged rollout. Each staged rollout is carried out for one feature, such that if a regression is detected, the feature development team can associate the impact back to the feature. But multiple features can be rolled out simultaneously in parallel, each in a separate rollout experiment, using the multi-layering system \cite{tang2010overlapping}. Upon setting up the rollout, the feature engineer specifies the target population for the rollout (e.g. all active users in US using the new app version 2.0 on Android), which is defined as all users. In this example, the feature is gradually rolled out through five stages, and each stage takes one day. The control group users get the default experience with the feature turned off, and the treatment group users get the new experience with the feature turned on. The monitoring will be performed on the metric data collected from these two groups. The untreated group get the same experience as the control group, but the data are not used for analysis, and these users are treated as outside of the experiment. The reason for keeping equal sample sizes between the control group and the treatment group is to avoid Simpson's Paradox \cite{crook2009seven}. 
During the rollout process, key user engagement and app health metrics are monitored by the sequential test to be introduced in the following section. The metrics can be defined on different analytics unit levels: trip, session, or user. And the metrics can be proportional metrics, continuous metric, and ratio metrics. For simplicity, the session level proportion metrics are used as examples. 

\begin {table*}
\centering
\caption {Example Staged Rollout Template}
\label{tab:staged_rollout_template}
{\small
\begin{tabular}{ | p{4cm} | p{1.2cm} | p{1.2cm} | p{1.2cm} | p{1.2cm} | p{1.2cm} | }
% * <mandieliu2015@gmail.com> 2018-07-31T01:35:40.204Z:
%
% > { | p{4cm} | p{1.2cm} | p{1.2cm} | p{1.2cm} | p{1.2cm} | p{1.2cm} | }
%
% ^.
\hline
Feature Flag & \multicolumn{5}{|l|}{Feature X} \\
\hline
Target Population & \multicolumn{5}{|l|}{Users in US using App Version 2.0 on Android} \\
\hline
Stages & 1 & 2 & 3 & 4 & 5 \\
\hline
Time & Day 1 & Day 2 & Day 3 & Day 4 & Day 5 \\
\hline
Control Sample Size & $1\%$ & $5\%$ & $20\%$ & $50\%$ & $0\%$ \\
\hline
Treatment Sample Size & $1\%$ & $5\%$ & $20\%$ & $50\%$ & $100\%$ \\
\hline
Untreated Sample Size & $98\%$ & $90\%$ & $60\%$ & $0\%$ & $0\%$ \\
\hline
Metrics  & \multicolumn{5}{|l|}{$\%$ Sessions with Login Success, $\%$ Sessions with Orders,} \\
& \multicolumn{5}{|l|}{$\%$ Sessions with App Crash, etc.} \\
\hline
\end{tabular}
}
\end{table*}

\section{Monitoring Algorithm - Sequential Test} \label{sec:monitoring}
In this section, a variant of the sequential test is introduced, called a mixture sequential probability ratio test (mSPRT), used as the core monitoring algorithm. In addition, a nonparametric variance estimation method (delete-a-group jackknife) is introduced to correct the correlation for the sequential algorithm. The statistical power and sample size estimation is derived for the sequential test as well.

\subsection{Mixture Sequential Probability Ratio Test (mSPRT)}
Sequential probability ratio test \cite{ghosh1991handbook} is widely used in clinical research, where scientists often allow sample size dependent decisions to be made based on likelihood ratio of two hypothesis. The mSPRT introduced in \cite{johari2017peeking} and \cite{pekelis2015new} applies to A/B testing to enable multiple tests without inflating the false positive rate (FPR). 
In an A/B testing setting, assume the control variables $\{X^{(i)}_{ctrl} \}_{i=1}^{n_{ctrl}}$ are independent random variables from a distribution with a density function $f(x|\mu_{ctrl}, \sigma_{ctrl})$ where $\mu_{ctrl}$ and $\sigma_{ctrl}$ represent mean and standard deviation. Similarly, the distribution density function for the treatment variables $\{X^{(i)}_{trt} \}_{i=1}^{n_{trt}}$ is $f(x|\mu_{trt}, \sigma_{trt})$. The hypothesis to be tested is on the difference in distribution mean: 
\begin{eqnarray}
H_{0}: & \delta := \mu_{trt} - \mu_{ctrl} = \delta_{0} \\
H_{1}: & \delta := \mu_{trt} - \mu_{ctrl} \neq \delta_{0}
\end{eqnarray}
where $\delta$ represents the difference in mean between treatment and control, and $\delta_0$ is the difference value under null hypothesis ($\delta_0=0$ for testing if two groups have the same metric mean). 
The test statistic used in mSPRT is a likelihood ratio integrated over a prior distribution of $\delta$ values under the alternative hypothesis. Denote the prior density function as $h(\delta)$, and for simplicity, a normal prior is chosen $h(\delta) \sim N(\delta_{0}, \tau)$ in this paper. It can be proven that the integrated likelihood ratio statistic is a Martingale under alternative hypothesis. The $(1-\alpha)$ confidence interval for $\delta$ is derived by \cite{pekelis2015new} as  
\begin{eqnarray}
\bar{x}_{trt} - \bar{x}_{ctrl} \pm \sqrt{
\frac{V(V + \tau)}{\tau}(-2log(\alpha) - log(\frac{V}{V+\tau})} 
\end{eqnarray}
with $\bar{x}_{trt}$ and $\bar{x}_{ctrl}$ as sample means and V as the variance of sample mean difference (e.g. $V = \frac{v_{ctrl}}{n_{ctrl}} + \frac{v_{trt}}{n_{trt}}$ where $v_{ctrl}$ and $v_{trt}$ are the sample variance estimates for control and treatment).

\subsection{Variance Estimation}
One assumption made by the sequential testing is the independence of each observation. However, this assumption often does not hold in practice. For example, if the click through rate is the metric of interest, it is improper to assume each impression is independent, since the same user can use the product multiple times on different days, and multiple impressions and clicks can be generated by the same user. Such observations are correlated since they are generated by the same user. Violating the independence assumption in the sequential test can produce an inflated false positive rate. Embedding a variance estimation with correlation correction is one way to generalize the mSPRT to correlated data.

Several previous papers discuss the adjustment of metric variance in A/B testing. The delta method \cite{deng2016data} and Bootstrap method \cite{bakshy2013uncertainty} are two variance estimation approaches that can be applied to correct the variance without the assumption of independence. These two methods work well in theory, however, they require storing raw data (e.g. all mobile event level data) in order to perform analysis. Therefore, they are hard to scale when the data storage is limited. In addition, processing the raw event level data adds computation latency which is a concern for monitoring all feature rollouts in real time. Therefore, a fast and scalable variance method is preferred. In our system, a version of the delete-a-group jackknife \cite{kott2001delete} variance estimation method is implemented, which meets such requirements. 

To implement the delete-a-group jackknife method, the users (user ID or device ID, which is the experiment unit) are split into $R$ partitions with equal probability using a hash function within each experiment group. The hash function (\cite{Kohavi2007KDD}, \cite{zhao2016online}) takes the user ID as input and outputs an integer as the partition ID $\in \{1,2,...,R \}$. Note that the hash function should be different (by choosing a different random seed or algorithm) from the one used for experiment randomization on splitting users into control and treatment groups, so that the user partition is independent to the experiment randomization. 

The delete-a-group jackknife variance estimator can be expressed as
\begin{eqnarray}
Var(\bar{X}) = \frac{R-1}{R} \sum_{r=1}^R (\bar{X}_{(r)} - \bar{X})^{2}
\end{eqnarray}

with 
\begin{eqnarray*}
\bar{X}_{(r)} =  \frac{1}{\sum_{i = 1}^n I(hash(uuid_i) \neq r)} \sum_{\{i: hash(uuid_i) \neq r\}} X_i
\end{eqnarray*}
where $\bar{X}_{(r)}$ is the metric mean for all users except partition $r$, $I(\cdot)$ is an indicator function with value $1$ if the argument condition is true and value $0$ otherwise, $hash(\cdot)$ is a hash function, and $uuid_i$ is the user ID for the $i$th observation. 

The delete-a-group jackknife variance estimation is scalable since only the partition level metrics are needed and stored for calculation, instead of the raw event-level data. Taking the CTR metric as an example, at each stage, the system will calculate the total number of impressions and clicks by partition, and then the cumulative (from the initial stage up to the current stage) CTR (as $\bar{X}_{(r)}$) can be calculated by dividing the sum of clicks by the sum of
impressions across stages except partition $r$. The overall cumulative CTR (as $\bar{X}$) can be calculated by dividing the sum of clicks by the sum of impressions, where both sums are across stages and partitions. 

\begin{figure}[ht]
\includegraphics[scale=0.1]{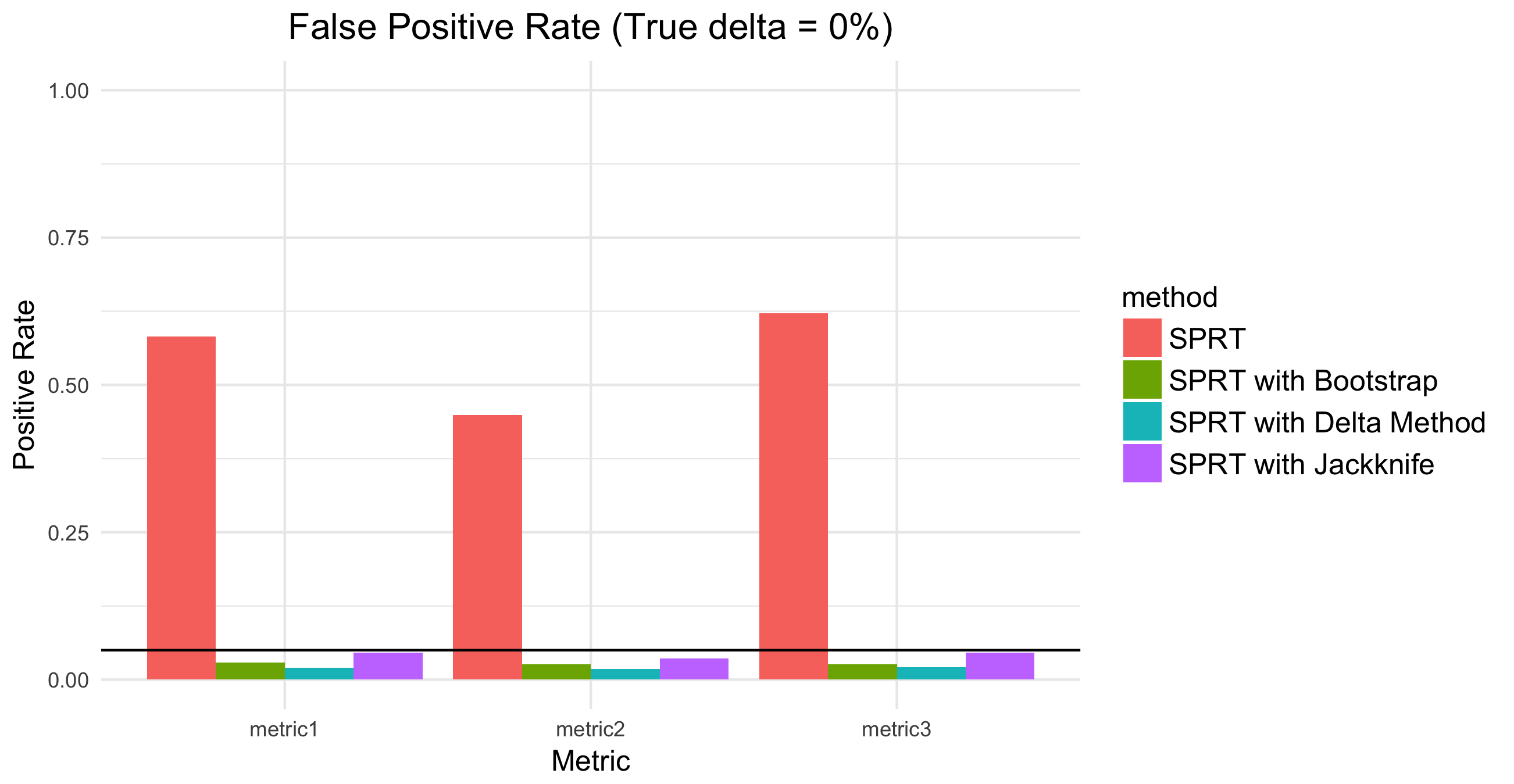}
\caption{Empirical False Positive Rate based on Different Variance Estimations}
\label{fig:var_fpr}
\end{figure}

\begin{figure}[ht]
\includegraphics[scale=0.1]{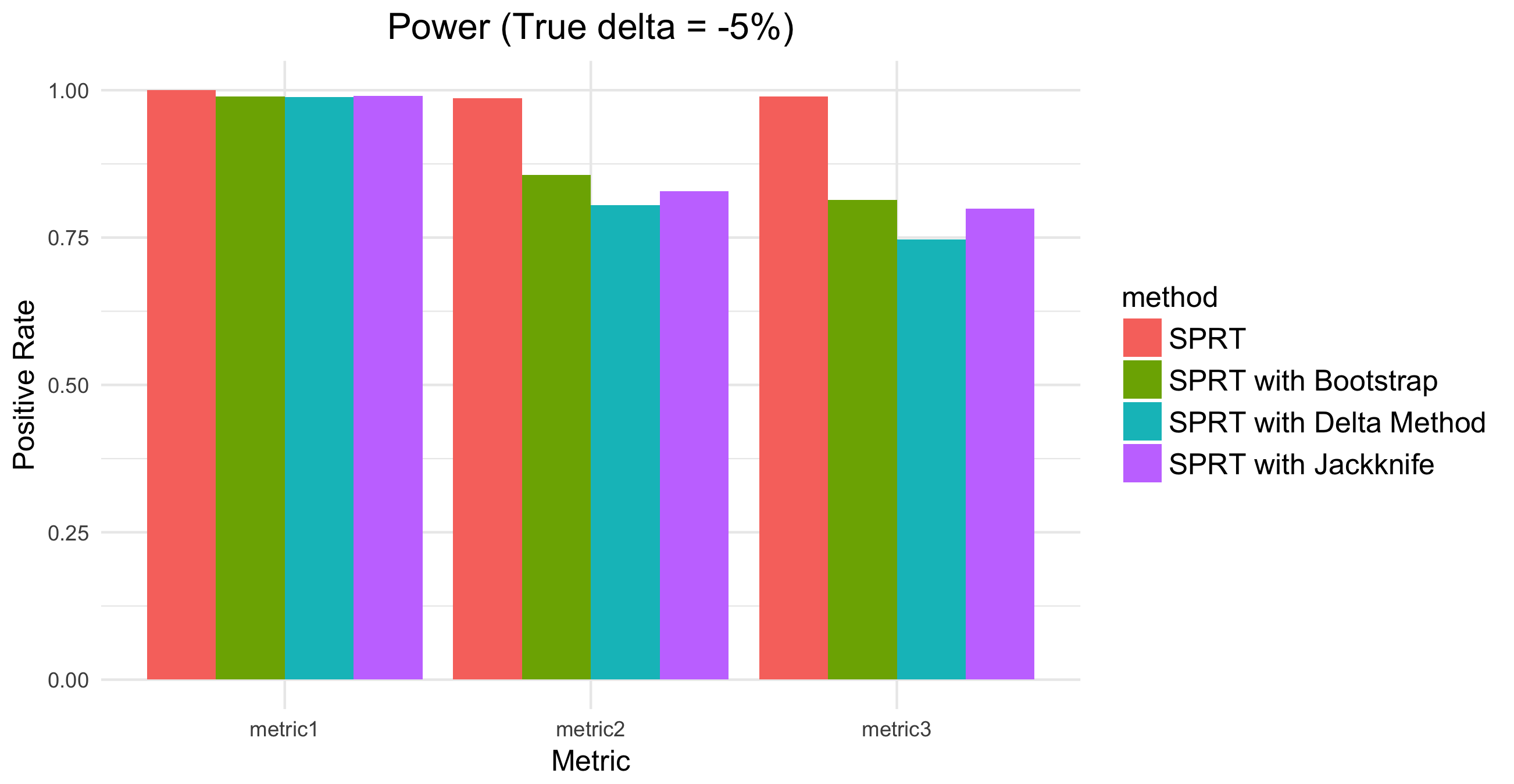}
\caption{Empirical Power based on Different Variance Estimations}
\label{fig:var_power}
\end{figure}

The empirical performance of different variance estimation methods are compared with real data reflecting the correlation structure in practice. Four methods are compared: sequential test (assuming observation independence), sequential test with bootstrap, sequential test with delta method, and sequential test with (delete-a-group) jackknife (with $R=10$). A sample dataset is prepared for three metrics in one-week time period. The false positive rate in Figure \ref{fig:var_fpr} is calculated by simulating an A/A test scenario: randomly assign the sample data into control and treatment groups by hashing the user ID. The power in Figure \ref{fig:var_power} is calculated by simulating an A/B test scenario: using the same experiment assignment logic as an A/A test, but artificially adding a $5\%$ metric increase in the treatment group. The data is fed into the algorithm at hourly level, and a positive result is defined if the test returns at least one significance (at $95 \%$ level) out of the $24 \times 7$ tests. The results show when the independence assumption does not hold, it leads to an inflated false positive rate for the original sequential test. All the three improved variance estimation methods successfully control the false positive rate below $5 \%$, while having a similar level of power. The jackknife method is preferred considering the scalability, and the performance can be improved by increasing the number of partitions $R$. By default, the jackknife method is applied for all examples related to real data in this paper.

\subsection{Power and Sample Size Estimation}
The statistical power of the the sequential test is needed for evaluating when it is ready to fully ramp up and conclude the rollout. At the beginning of the rollout, the rollout developer will set a sensitivity level for metric degradation detection. For example, the developer can set the system to raise an alert if the true metric difference is greater than $2\%$. In this paper, this minimum detectable difference is defined as the MDE (minimum detectable effect). Then the criteria for the feature to be formally rolled out is to achieve enough statistical power to detect the MDE. If the power is achieved, and there is no regression detected, then system will believe this feature is not causing a regression greater than the MDE with high confidence. In order to make the power-based criteria work, an estimation of power and sample size is needed during the rollout process. 

To derive the power and sample size estimation formula, define the stopping time $N_{\delta}$ as the smallest sample size that the sequential test becomes significant given the true difference as $\delta$. Note $N_{\delta}$ is a random variable. The power of the sequential test for a finite sample size can be expressed as the probability of the sequential process becoming significant before the given sample size:
\begin{eqnarray}
Pr(N_{\delta} < N_{{\delta}_{1-\beta}} | {\delta}) = 1 - \beta
\end{eqnarray}
where $1 - \beta$ indicates the power and $N_{{\delta}_{1 - \beta}}$ denotes the sample size threshold for achieving the power $1 - \beta$ given $\delta$.

Following the intuition that $N_{{\delta}_{1 - \beta}}$ is the $1 - \beta$ percentile of the distribution for $N_{\delta}$, it can be approximated by a linear combination of the distribution mean and standard deviation:
\begin{eqnarray}
N_{{\delta}_{1 - \beta}} \approx E[N_{\delta}] + f(\beta) * \sqrt{(Var(N_{\delta}))}
\end{eqnarray}

A previous Monte Carlo study \cite{cox1966note} found that the variance of the sample size needed to detect a given MDE for sequential test is approximately proportional to the square of the average sample size: $Var(N_{\delta}) \propto (E[N_{\delta}])^2$. 

In addition, the average sample size can be approximated by $E(N_{\delta}) \approx \frac{v_{x} + v_{y}}{{\delta}^{2}}\{log(-2 log \alpha) - 2 log \alpha \}$ by taking $\tau = {\delta}^{2}$ (see Appendix). 

To estimate the coefficient $f(\beta)$, an empirical power curve is fitted in a simulation study that yields the following approximation for the sample size required to achieve a given statistical power: 
\begin{eqnarray}
N_{{\delta}_{1-\beta}} &\approx & E(N_{\delta}) + f(\beta) E(N_{\delta}) \\ 
&\approx& (0.35 - 0.79 log\beta)(\frac{v_{x} + v_{y}}{{\delta}^{2}}) \\
&& \times (log(-2 log \alpha) - 2 log \alpha)
\end{eqnarray}

\begin{figure}[ht]
\includegraphics[scale=0.6]{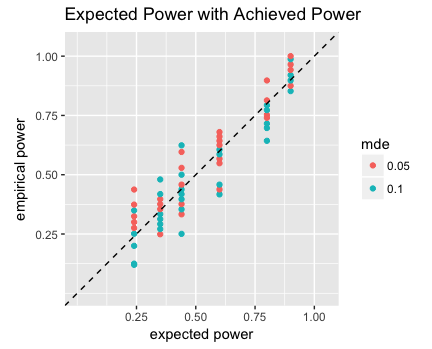}
\caption{Evaluation of Power Estimation. (The Y-axis shows the empirical power in simulation, and the X-axis shows the expected power calculated by the power function. The solid line is the diagonal line illustrating the good of fit of the power function.)}
\label{fig:est_power}
\end{figure}

Given this formula, the power can be estimated by solving the formula for $\beta$ for a given sample size $N_{{\delta}_{1-\beta}}$. This power estimation works well empirically for sample size larger than $500$. Figure~\ref{fig:est_power} shows the estimated power compared with the actual power.

\section{Ramp-up Algorithms}\label{sec:rampup}
During the rollout, when the monitoring algorithm raises an alert, the rollout will be paused or reverted for investigation. However, if the monitoring algorithm does not raise an alert, it does not mean the feature is guaranteed to be flawless, but it could be the case that the sample size is not sufficient for the monitoring algorithm to catch the regression at certain levels of sensitivity. With the uncertainty in the feature performance, rolling out too fast could expose more than necessary users to the feature. Therefore, a separate ramp-up algorithm is needed to decide how to increase the rollout percentage from one stage to the next stage. 
Next, three approaches for ramping up the rollout are introduced. 

\subsection{Time-based Ramp-up} \label{sec:power-based} 
The time-based ramp-up schedule requires engineers to decide the stages and corresponding sample size percentage before the rollout begins. In addition, the time between each step is specified. After the rollout starts, if there is no alert made by the monitoring algorithm for the given time length, the rollout will ramp up to the next stage. In the example shown in Table~\ref{tab:staged_rollout_template}, if the rollout is set to be ramping up every day, then it will be fully rolled out in five days if there is regression captured.
The advantage of the time-based ramp up is to give the developer more control on the rollout. However, since this ramp-up schedule does not adapt to the information collected during the rollout, it may suffer in 1) ramping up too quickly when the risk and uncertainty is high or 2) ramping up too slow when the data shows high level of certainty.

\subsection{Power-based Ramp-up} \label{sec:power-based}
The power-based ramp-up schedule aims to achieve sufficient statistical power for detecting the pre-defined MDE $\delta_{MDE}$ within a given time frame. For example, if the feature developer wants to roll out a feature with $80\%$ power to detect potential $2\%$ metric difference within one week, then the power-based ramp-up schedule will adjust the sample size adaptively to achieve the rollout goal. 
At each stage, if the observed difference $\hat{\delta}$ is less than pre-defined MDE $\delta_{MDE}$, then the sample size required for the MDE is calculated. Otherwise, if the observed difference is bigger than MDE, then the sample size required for detecting the current difference is calculated. This way, the system makes sure the ramp-up does not put too many users under risk if the observed signal is negative. In addition, a maximum ramp-up proportion threshold can be put for each stage, so the ramp-up speed is not too aggressive. In sum, the sample size recommended by the power-based algorithm is
$$min(N_{\delta_{MDE,1-\beta}}, N_{\hat{\delta}_{1-\beta}}, N_{stage\_limit})$$
where $N_{stage\_limit}$ is the maximum sample size allowed at a given stage, $N_{\delta_{1-\beta}}$ is the sample size required for the MDE $\delta$, and $N_{\hat{\delta}_{1-\beta}}$ is the sample size corresponding to the observed difference $\hat{\delta}$. The recommended sample size can then be transformed into sample size percentage by taking account of the predicted total  sample size for the next stage.

%%\begin{figure*}[ht]
%%\includegraphics[scale=0.38]{power_based_2018.png}
%%\caption{Estimated Power vs. Actual Power}
%%\label{fig:power_rampup}
%%\end{figure*}

\subsection{Risk-based Ramp-up} \label{sec:risk-based}
Risk-based ramp-up algorithm proceeds the rollout at the maximum allowed speed on a tolerable risk level. For example, if the risk criteria is set as "avoiding loosing trips from more than 1000 users". Then at any given stage, the system can calculate what the maximum number of extra users that could be exposed given the current experiment signals. 

To state this idea in formal formulation, the potential risk of uncertain metric degradation is quantified and controlled by a probability threshold.
Without loss of generality, assume the metric of interest is a proportion metric, e.g. $\%$ users with an order (the continuous metric scenario can be derived in a similar approach). The negative impact of the feature can be quantified as number of treatment users not making an order because of the feature: $N_{trt}(\mu_{trt} - \mu_{ctrl}) = N_{trt} \delta$. The risk tolerance during the rollout can be defined as:
\begin{eqnarray}
\label{risk_formula}
Pr(N_{trt} \delta \leq -C | D) \leq R
\end{eqnarray}
where 
\begin{itemize}
\item $C$ is a positive constant, as the pre-specified tolerable cost threshold. One choice can be $C = 
\delta_{MDE} * N_{\delta_{MDE,1-\beta}}$, that allows enough cost budget to yield sufficient statistical power.
\item $R$ is the pre-specified risk probability threshold. $R$ is a parameter to trade off the rollout safety and speed.
\item $D$ indicates data, which means the probability specified is based on a posterior distribution given the data.
\end{itemize}

Note that the formula above is an example of one-sided test assuming only potential decrease in the metric is a risk to be controlled. This framework can be easily generalized to one-sided increase risk and two-sided risk specification. Without loss of generality, the one-sided decrease risk is used in the following derivation for simplicity.

In order to solve the inequality (\ref{risk_formula}), the posterior distribution $p(\delta | D)$ can be derived as follows. For simplicity, assuming control and treatment have equal sample size and equal variance. By the central limit theorem, assume the sample mean follows a normal distribution $\bar{x}_{ctrl} \sim  N(\mu_{ctrl}, \sigma^2/n)$ and 
$\bar{x}_{trt} \sim  N(\mu_{trt}, \sigma^2/n)$,
where $\mu_{ctrl}$ and $\mu_{trt}$ are the distribution means, $\sigma^2$ is the common variance, and $n$ is the sample size. In addition, assume the prior distributions for the mean parameter $\mu_{ctrl}$ and $\mu_{trt}$ are 
$q(\mu_{ctrl}) \sim  N(\mu_0, \sigma_0^2)$ and $q(\mu_{trt}) \sim  N(\mu_0+\delta_0, \sigma_0^2)$, 
where $\mu_0$, $\sigma_0^2$, and $\delta_0$ are prior distribution parameters, indicating prior control mean, prior common variance, and prior difference parameter. Under these assumptions, the posterior distribution can be derived as a normal distribution (see \cite{bolstad2016introduction}): 
\begin{eqnarray}
p(\delta | D) \sim N(m_{\delta}, s^2_{\delta})
\end{eqnarray}
where 
\begin{eqnarray*}
m_{\delta} &=& \frac{1/\sigma_0^2}{n/\sigma^2 + 1/\sigma_0^2} \delta_0 + \frac{n/\sigma^2}{n/\sigma^2 + 1/\sigma_0^2} (\bar{x}_{trt} - \bar{x}_{ctrl}) \\
s^2_{\delta} &=& \frac{2 \sigma^2 \sigma_0^2}{\sigma^2 + n\sigma_0^2}
\end{eqnarray*}
Note that the only unknown parameter in this posterior distribution is the metric variance $\sigma^2$. In practice, this parameter can be estimated and substituted by the pooled sample variance.  

At stage $t$, the observed cumulative sample size for treatment is $\sum_{s=1}^t n_{trt, s}$. The risk-based ramp-up algorithm aims to decide an appropriate sample size for next stage $n^*_{trt, t+1}$ that controls the potential risk. The risk inequality (\ref{risk_formula}) can be written as
\begin{eqnarray}
& Pr(N_{trt} \delta \leq -C | D) \\
= & Pr((\sum_{s=1}^t n_{trt, s} + n^*_{trt, t+1})  \delta \leq -C | D) \\
= & Pr(\delta \leq \frac{-C}{(\sum_{s=1}^t n_{trt, s} + n^*_{trt, t+1})} |D) \\
\leq & R
\end{eqnarray}

The left side of the inequality becomes the cumulative distribution function of the posterior distribution of $\delta$. The the maximum tolerable sample size $n^*_{trt, t+1}$ can be derived as:

\begin{eqnarray}
\frac{1}{s_{\delta}} \cdot (\frac{-C}{(\sum_{s=1}^t n_{trt, s} + n^*_{trt, t+1}) } - m_{\delta}) = \Phi^{-1}(R) \\ 
n^*_{trt, t+1} =  \max(\frac{- C}{s_{\delta} \Phi^{-1}(R) + m_{\delta}} -  \sum_{s=1}^t n_{trt, s}, 0)
\end{eqnarray}
The corresponding treatment rollout percentage for the next stage $p_{t+1}$ can be calculated by dividing the treatment rollout sample size $n^*_{trt, t+1}$ by a predicted total population size for the next stage $\hat{n}_{total, t+1}$ as:
$$p_{t+1} = \max( \min ( p_t, \frac{n^*_{trt, t+1}}{\hat{n}_{total, t+1}}), 0.5)$$

The $\min$ and $\max$ conditions are added to ensure the rollout percentage is monotonic increasing (unless reverting to $0$), and the maximum rollout percentage that can be achieved by the ramp-up algorithm is $50\%$. The final decision from ramping up from $50\%$ to $100\%$ (in one stage) is determined by power calculation, which is beyond the control of the ramp-up algorithm. The predicted total population size for the next stage $\hat{n}_{total, t+1}$ can be estimated by a generic time series model. The discussion of this prediction model is beyond the scope of this paper. 

\section{Examples} 
\label{sec:example}

\begin {table*}
\centering
\caption {Empirical Performance of Ramp-up Algorithms based on Synthetic Data under A/A and A/B Tests (NA is shown when the observed cases are too few to draw a conclusion.)}
\label{tab:staged_rollout_performance_simulated_data}
{\small
\begin{tabular}{ | p{2.5cm} | p{1.5cm} | p{1.5cm} | p{1.5cm} |p{1.5cm} | p{1.5cm} | p{1.5cm} | p{1.5cm} | p{1.5cm} |}
\hline
Metrics & Time-based A/A (Low Speed)& Time-based A/B (Low Speed)& Time-based A/A (High peed)& Time-based A/B (High Speed)& Power-based A/A & Power-based A/B & Risk-based A/A & Risk-based A/B \\
\hline
Positive Rate (FPR for AA, Power for AB) & 0.04& 0.985 & 0.015 & 0.985 & 0.015 & 0.975 & 0.01 & 0.98 \\
\hline
Average Time before Detection (h) & NA & 63 & NA & 55 & NA & 57  & NA & 52 \\
\hline
Average Time before Fully Rollout (h) & 61 & NA & 57 & NA & 52 & NA & 47 & NA \\
\hline
Weighted Average Rollout Pct before Detection & NA & $1\%$,$5\%$, $17\%$,$14\%$ & NA & $1\%$,$10\%$, $20\%$,$16\%$ & NA & $1\%$,$13\%$, $17\%$,$9\%$, $4\%$,$1\%$ & NA & $1\%$,$18.3\%$, $18.5\%$,$10\%$, $3\%$,$1\%$ \\
\hline
Weighted Average Rollout Pct before Fully Rollout & $1\%$,$5\%$, $31\%$,$87\%$ & NA & $1\%$,$10\%$, $39\%$,$91\%$ & NA & $1\%$,$17\%$, $25\%$,$65\%$, $94\%$,$99\%$ & NA & $1\%$,$17\%$, $34\%$,$73\%$ & NA \\
\hline
Average Sample Size Used before Detection & NA & 12085 & NA & 11803 & NA & 11528 & NA & 11857 \\
\hline
Average Sample Size Used before Fully Rollout & 9591 & NA & 10186 & NA & 9876 & NA & 9300 & NA \\ 
\hline
Avg Total Loss & 21 & 136 & 18 & 137 & 20 & 132 & 18 & 136 \\
\hline 
$\%$ of Tests Exceeding the Loss Tolerance $C$ & $0\%$ & $5\%$ & $0\%$  & $9\%$ & $0\%$ & $7\%$ & $0.05\%$ & $10.05\%$ \\
\hline
\end{tabular}
}
\end{table*}

\begin {table*}
\centering
\caption {Empirical Performance of Ramp-up Algorithms based on Real Data under A/A and A/B Tests (NA is shown when the observed cases are too few to draw a conclusion.)}
\label{tab:staged_rollout_performance_real_data}
{\small
\begin{tabular}{ | p{2.5cm} | p{1.5cm} | p{1.5cm} | p{1.5cm} |p{1.5cm} | p{1.5cm} | p{1.5cm} | p{1.5cm} | p{1.5cm} |}
\hline
Metrics & Time-based A/A (Low Speed)& Time-based A/B (Low Speed)& Time-based A/A (High peed)& Time-based A/B (High Speed)& Power-based A/A & Power-based A/B & Risk-based A/A & Risk-based A/B \\
\hline
Positive Rate (FPR for AA, Power for AB) & 0.01 &0.97 & 0.01 & 0.99 & 0.015 & 0.99 & 0.01 & 0.99 \\
\hline
Average Time before Detection (h) & NA &58 & NA &52 & NA &54 & NA &44 \\
\hline
Average Time before Fully Rollout (h) &67 &NA & 63 & NA& 60 &NA & 52 & NA \\
\hline
Weighted Average Rollout Pct before Detection &NA &$1\%$,$5\%$, $20\%$,$25\%$ & NA &$1\%$,$10\%$, $20\%$,$25\%$ & NA & $1\%$,$12\%$, $15\%$,$12\%$, $7\%$,$3\%$, $1\%$  & NA & $1\%$,$18\%$, $19\%$,$7\%$, $1\%$\\
\hline
Weighted Average Rollout Pct before Fully Rollout &$1\%$,$5\%$, $24\%$,$77\%$ &NA &$1\%$,$10\%$, $25\%$,$86\%$ &NA & $1\%$,$16\%$, $20\%$,$56\%$, $87\%$,$97\%$ &  NA & $1\%$,$16\%$, $42\%$,$55\%$, $98\%$ &NA \\
\hline
Average Sample Size Used before Detection &NA &11126 &NA &10981 & NA &9282 & NA & 9438 \\
\hline
Average Sample Size Used before Fully Rollout &13281 &NA &13104 &NA & 13420 &NA & 12420 & NA \\
\hline
Avg Total Loss &30 &135 &27 & 141 & 29 &124 & 20 & 127 \\
\hline
$\%$ of Tests Exceeding the Loss Tolerance $C$ &$0\%$ & $12\%$ & $0\%$ & $22\%$ & $0\%$ & $14\%$ & $0\%$ & $12\%$ \\
\hline
%\hline
\end{tabular}
}
\end{table*}

This section evaluates the empirical performance of staged rollout framework with the three ramp-up algorithms on both real data and synthetic data examples. 

The real data comes from a feature rollout experiment, which does not cause a regression. One of the key session level metric monitored is a binary metric with mean $0.7$. However, the session level metric can not be regarded as i.i.d. sample from Bernoulli distribution, since the same user can have multiple sessions. The data is collected hourly, and there are $7$ days' data in total. On the other hand, the synthetic data is generated as i.i.d. sample from a Bernoulli distribution with mean $0.7$. 
During the rollout, the sequential test runs every hour and the rollout is reverted when a significant metric degradation is detected. The criteria for rolling out to all users is to achieve a MDE of $0.05$ at power of $0.9$. 
If no regression is detected, the system ramps up the rollout to the next stage on the next day. Although the ramp-up frequency is fixed (daily), the sample size for the next stage is determined by the three ramp-up algorithms. 

Under this setting, the three ramp-up algorithms are evaluated through both A/A and A/B tests, each with $200$ replications. 
Each trial of A/A test is generated by randomly splitting the data into treatment and control groups by hashing user ID. The success of an A/A test can be defined as the feature getting fully rolled out in a fast speed. 
Each trial of A/B test is first generated by the same random split as A/A test, but then an artificial metric difference is introduced to the treatment group. For each metric value in the treatment group, a relative difference is drawn from gamma distribution with shape $= 6.25$ and scale $= 0.008$ (indicating mean as $5\%$ and standard deviation as $2\%$), that is applied to create the artificial difference. The success of an A/B test can be defined as detecting the difference with relatively small sample size.  

The Time-based ramp-up algorithm is tested at two speed levels with rollout percentages: ($1\%$, $5\%$, $20\%$, $50\%$) and ($1\%$, $10\%$, $20\%$, $50\%$). The Power-based ramp-up algorithm sets the target MDE as $5\%$. For the risk-based ramp-up algorithms, we set prior parameters as $\delta_0 = 0.05$, prior variance $\sigma_0^2 = 0.0004$, and risk probability threshold $R = 0.1$. Note that a non-zero prior mean is set in order to be conservative about the feature performance in the beginning. In this example, we are controlling the risk in the increasing direction. 

Table \ref{tab:staged_rollout_performance_simulated_data} and Table \ref{tab:staged_rollout_performance_real_data} present the results of different rollout approaches. Overall, all the three algorithms achieve false positive rate below $0.05$ for A/A tests, and power above $0.9$ for A/B tests. 
The Time-based algorithm, as a simple benchmark approach, seems to take longer time and more samples to make a final correct decision. 
The Power-based ramp up algorithm gives  relatively stable performance. Since this algorithm is set up based on the sequential test used for monitoring, it often uses the smallest sample size to detect the regression. This is reasonable since the power-based algorithm estimates the number of sample size needed for the regression detection, which prevents over-exposing users.  
The Risk-based algorithm on average achieves the earliest final detection in both A/A and A/B test. It also controls the risk to be around the desired level $R = 10\%$ ($10.05\%$ in simulated data example and $12\%$ in real data example).  In contrast, both time-based and power-based ramp-up algorithms do not exhibit consistent risk control performance. 

Note that there are parameters that can be tuned for each of the three algorithms, especially the risk-based algorithm have multiple parameters. The results only display the outputs based on specific parameters. The performance order can change with different parameter settings and preference over safety and speed. In practice, such parameters can be determined or tuned through simulation before the actual rollout begins.  

\begin{figure}[ht]
\includegraphics[scale=0.16]{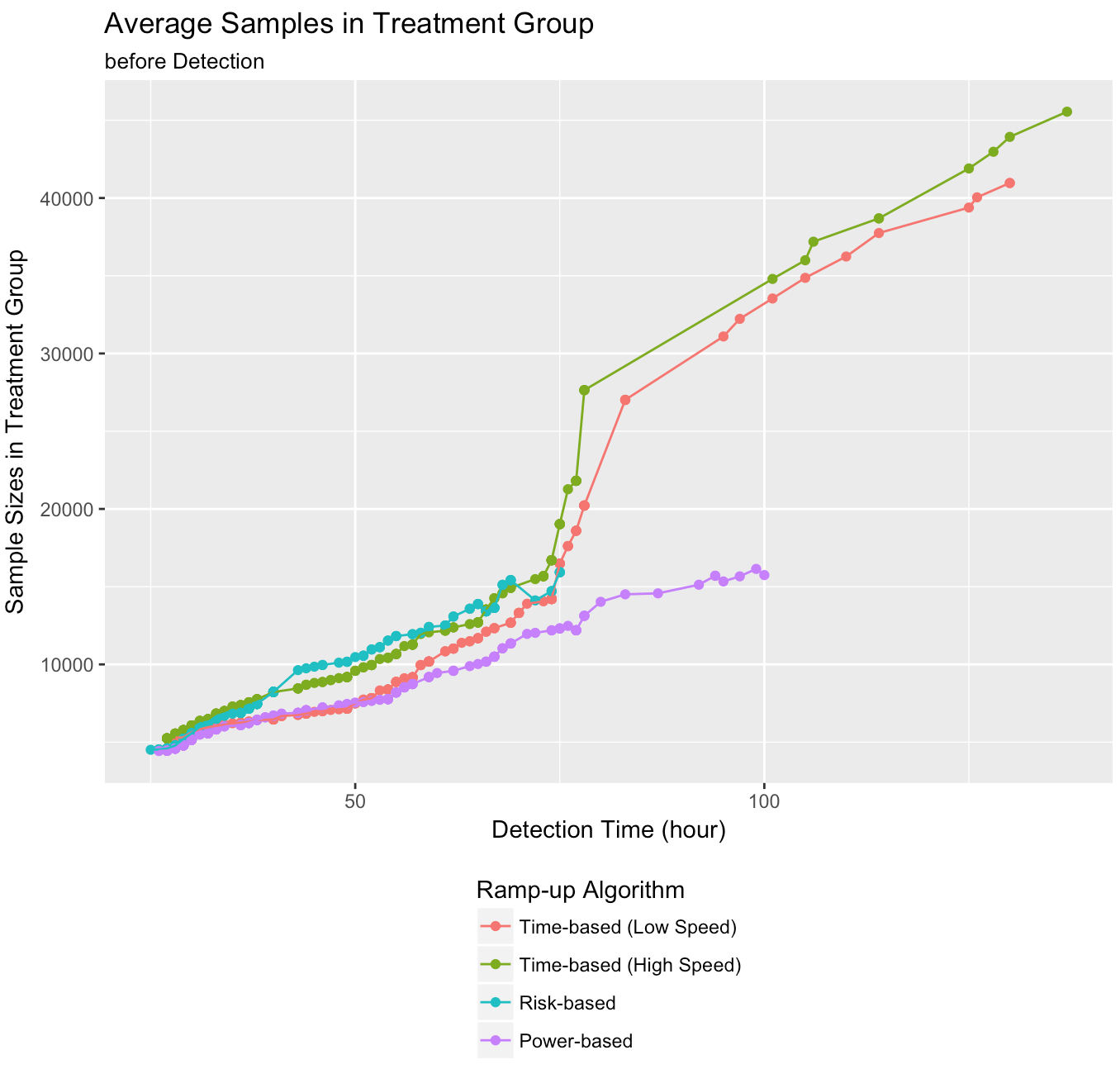}
\caption{Average Sample Size Exposed before Detection under A/B Test on Real Data}
\label{fig:compare_rampup}
\end{figure}

\section{Conclusion} \label{sec:discussion}
In this paper, a staged rollout framework is presented to automatically release a feature with special attention to optimizing the safety and speed. This framework can be used broadly for code changes to product, both with and without an A/B experiment. In our experiments, the variance corrected sequential test with sample size and power calculation accommodates the practical needs as a continuous monitoring solution. The three ramp-up algorithms represent different perspectives of trade-offs made between safety and speed. 

The empirical evaluation of the system shows the sequential monitoring system can control false positive rate effectively while deliver a reasonable power. All three ramp-up algorithms, when configured properly, can make reasonable trade-offs between safety and speed with emphasis on different aspects. This framework has been tested and proven useful in practice for regression detection and ramp-up management to assist developers with rolling out both safely and quickly. 

For implementing the ramp-up algorithms in production, a practical recommendation is to start with time-based algorithm, and then power-based algorithm or risk-based algorithm. In practice, the time-based ramp-up algorithm is easy to implement and it also gives more control and transparency to the feature development team. Also compared with a direct rollout, the time-based algorithm can provide sufficient safety for most business case scenarios. If lack of statistical power is a concern, then the power-based algorithm can be further used. Otherwise, if the risk during the rollout period needs to be further controlled, the risk-based algorithm can be utilized.

While the examples of this paper mainly focus on one metric case for illustration, the staged rollout framework can adapt to the case with multiple metrics. The monitoring algorithm can monitor multiple metrics in parallel, and the significance level can be adjusted to control for family-wise false positives. While a large set of metrics can be put under the monitoring system, a smaller set of key metrics can be selected to be used in the ramp-up algorithm. The time-based ramp-up can be extended to multi-metric case naturally. The power-based ramp-up can take the maximum over the recommended sample sizes across different metrics to achieve the power. The risk-based ramp-up can take the minimum over the recommended sample sizes across different metrics to control risk. 

Another practical consideration in the rollout system is to monitor the data quality of the rollout. Poor data quality (caused by logging system, outliers, etc.) can lead to inflation of false positives or false negatives \cite{zhao2016online}. Data quality check \cite{appiktala2017demystifying, chen2017faster} and outlier removal \cite{he2017probabilistic} is useful in practice for the rollout system. When a red flag is raised on data quality, the rollout platform team and the feature development team need to engage and investigate the root cause. Detailed discussion on the data quality monitoring and diagnosis is out of the scope of this paper.  

The power-based ramp-up algorithm is built on the frequentist inference, and the risk-based ramp-up algorithm is based on Bayesian framework. One direction of future development is using reinforcement learning to guide the rollout process as a Markov Decision Process. Instead of focusing on statistical power and potential risk, the reinforcement learning algorithm will make the trade-off between the value of rolling out the new feature versus the cost associated with the rollout. It will eventually optimize the final reward given the value and cost specification of the feature. This algorithm is currently in development.

\section{Appendix}
\subsection{Prior Distribution for mSPRT}
In Section \ref{sec:monitoring}, the mSPRT test statistic is presented with a prior distribution $h(\delta) \sim N(0, \tau)$. 
Here we describe a practical choice for parameter $\tau$. Although choice of $\tau$ value does not affect the martingale and the statistical test property in theory, it affects the empirical convergence speed (i.e. sample size needed to achieve significance) in practice. 
Pollak and Siegmund \cite{pollak1975approximations} derived the average sample size needed to detect the real difference $\delta$ under sequential analysis. In our context, the following formula provides a reasonable approximation to the average sample size: 
\begin{eqnarray}
E[N_{\delta}] \approx \frac{v_{ctrl} + v_{trt}}{{\delta}^{2}}\{log[\frac{-2\tau e^{{\delta}^{2}/{\tau}} log(\alpha)}{{\delta}^{2}\alpha^{2}}]-1\}.
\end{eqnarray}

\begin{figure}[ht]
\includegraphics[scale=0.6]{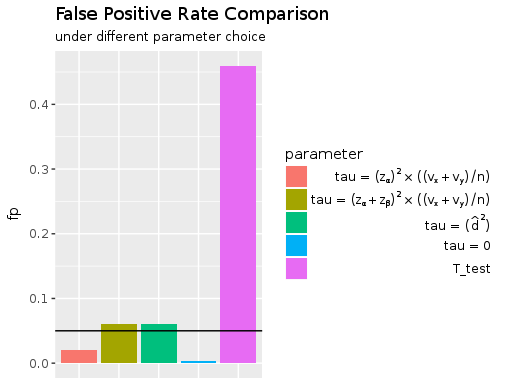}
\caption{False Positive Rate based on Different Priors}
\label{fig:fpr_compare}
\end{figure}

\begin{figure}[ht]
\includegraphics[scale=0.6]{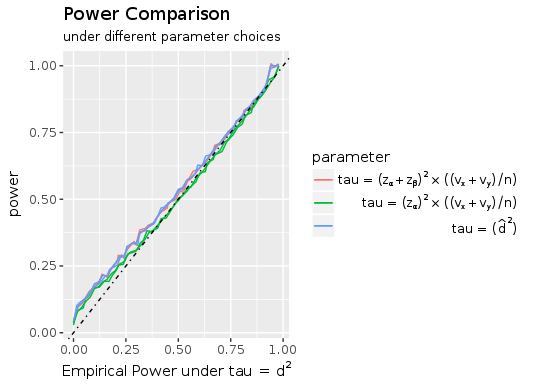}
\caption{Empirical Powers based on Different Priors (The X-axis shows the empirical power under $\tau = d^2$ and the Y-axis shows the empirical power achieved by alternative parameter settings.)}
\label{fig:pw_compare}
\end{figure}

Taking derivatives yields $\tau = {\delta}^{2}$ to achieve the smallest sample size. However, as the real difference $\delta$ is unknown to experimenters, we want to find a substitute for $\delta$ to provide comparable type I and type II error. 
Compared with fixed-horizon t test, where we have the sample size $n \propto \frac{v_{ctrl} + v_{trt}}{{\delta}^{2}}$. We want to use $\tau = {\delta}^{2} = A \frac{v_{ctrl} + v_{trt}}{n}.$ Below is a simulation illustrating the choice of $\tau$ to type I and type II error. \\
 A sample data is generated from binomial distribution with p varying from 0.1 to 0.9 and relative difference equal to 0 or 0.05. Along the sequence, 200 checks are conducted in equal interval with each checkpoint using all previous observations. Treatment and control group have equal sample size per each check. A positive result is defined as the test reports at least one significance (at $95 \%$ level) out of the 200 tests, which indicates false positive rate in Figure \ref{fig:fpr_compare} and power in Figure \ref{fig:pw_compare}

From above simulation, we see that using $\bar{\delta}^{2}$ i.e. the square of Observed Difference and using $\tau = (z_{1-\alpha / 2} + z_{1-\beta / 2})^{2} \frac{v_{ctrl} + v_{trt}}{n}$ has similar performance in both empirical power and false positive rate. This is because in fixed-horizon test, the average sample size under Type I error $= \alpha$ and Type II error $=\beta$ has the following relationship $ z_{1-\alpha / 2} + z_{1-\alpha / 2} = \frac{\delta}{\sqrt{(v_{ctrl} + v_{trt})/n}}$. In Figure \ref{fig:pw_compare}, the empirical power under three parameter choices are plotted against the empirical power under $\tau = {\delta}^{2}$. When the empirical power under $\tau = {\delta}^{2}$ is small, the other three choices all yield relatively bigger power which is because of an overestimate of $\tau$ due to small sample size. This discrepancy dies down as the empirical power of $\tau = {\delta}^{2}$ increases. To balance out the type I and the type II error, we chose $A = z_{1-\alpha/2}^{2}$, yielding $\tau = {\delta}^{2} = z_{1-\alpha / 2}^{2} \frac{v_{ctrl} + v_{trt}}{n}$ in practice.

%ACKNOWLEDGMENTS are optional
\section*{Acknowledgment}
We would like to extend our thanks to Akash Parikh, Donald Stayner, and Anando Sen for insightful discussion on defining the problem and forming the solution, to Sisil Mehta and Tim Knapik for thoughtful design and implementation of the algorithms in production. Special thanks to Professor Peter Dayan on discussion and formulation of Reinforcement Learning algorithm as future development. We would also like to thank Olivia Liao for her early research and empirical work of applying sequential test on the experimentation platform.

\bibliographystyle{./IEEEtran}
\bibliography{./IEEEabrv,./IEEEexample}

\end{document}